\documentclass[aps,superscriptaddress,showpacs,twocolumn]{revtex4}
\usepackage{graphicx}

\begin{document}

\title{Minimal tomography with entanglement witnesses}
\author{Huangjun Zhu}%
\affiliation{Centre for Quantum Technologies, %
National University of Singapore, Singapore 117543, Singapore}
\affiliation{NUS Graduate School for Integrative Sciences and
Engineering, Singapore 117597, Singapore}
\author{Yong Siah Teo}%
\affiliation{Centre for Quantum Technologies, %
National University of Singapore, Singapore 117543, Singapore}
\affiliation{Department of Physics, %
National University of Singapore, Singapore 117542, Singapore}

\author{Berthold-Georg Englert}
\affiliation{Centre for Quantum Technologies, %
National University of Singapore, Singapore 117543, Singapore}
\affiliation{Department of Physics, %
National University of Singapore, Singapore 117542, Singapore}

\begin{abstract}
We introduce informationally complete measurements whose outcomes are
entanglement witnesses and so answer the question of how many witnesses
need to be measured to decide whether an arbitrary state is entangled or not:
as many as the dimension of the state space.
The optimized witness-based measurement can provide exponential improvement
with respect to witness efficiency in high-dimensional Hilbert spaces,
at the price of a reduction in the tomographic efficiency.
We describe a systematic construction, and illustrate
the matter at the example of two qubits.
\end{abstract}

\date{22 June 2009}

\maketitle

Quantum entanglement is a useful resource with a wide range of
applications such as quantum teleportation
\cite{Bennett,Bouwmeester, Boschi}, quantum key distribution
\cite{Ekert}, and  quantum computation \cite{Jozsa}. Crucial for the
implementation of all these nonclassical tasks  are the detection
and characterization of entanglement in experiments. Entanglement
can be detected by measuring entanglement \emph{witnesses}
\cite{HorodeckiM, Bourennane}. How many witnesses do we have to
measure to determine whether a generic unknown quantum state is
entangled or separable? \cite{Ekert 2} If we were to rely on each
witness separately, we would have to measure an infinite number of
them even in the case of two qubits, as we will see below.

On the other hand, state tomography with an informationally complete
(IC) probability operator measurement (POM) \cite{Prugovecki, Busch,
D'Ariano,PO(V)M}, which may consist of no more than $d^2$ outcomes ($d$ is
the dimension of the Hilbert space, a \emph{minimal} IC-POM has
$d^2$ outcomes) can provide accurate information about an unknown
input state, if sufficiently many input states are available. With
this tomographic information at hand,  we can reconstruct the
unknown input state and then determine its  separability, possibly
exploiting well known criteria, such as the PPT criterion
\cite{Peres, HorodeckiM}, the range criterion \cite{HorodeckiP}, the
matrix realignment criterion \cite{Rudolph,Chen}, the covariance
matrix criterion \cite{Guhne}, or perhaps others, or making use of
available algorithms \cite{Lewenstein1,Hulpke}. If we can design a
minimal IC-POM such that every outcome is a witness, $d^2$ witnesses
will suffice to determine the separability of any unknown input
state.

We establish that such witness operator measurements (WOMs) do exist
and show how to construct WOMs with optimal witnesses
\cite{Lewenstein2} as outcomes from rank-one POMs --- POMs in which
each outcome is a subnormalized projector onto a pure state. These
WOMs can provide exponential improvement of the witness efficiency
at the price of a much lower tomographic efficiency. The trade-off
between witness efficiency and tomographic efficiency turns out to
be quite common, and it is impossible to construct a minimal IC-POM
which achieves both the best witness efficiency and optimal
tomographic efficiency.

We shall be concerned with a bipartite system whose parts have
subsystem dimensions $d_1$, $d_2$ with $d_1\leq d_2$, and the total
dimension is $d=d_1d_2$. Usually, a witness is a Hermitian operator
with nonnegative mean values for all separable states, and negative
mean values for some entangled states \cite{HorodeckiM}, but this
convention of the threshold at zero is not expedient here. For us, a
witness is a positive operator with mean values less than or equal
to some threshold for all separable states, and mean values above
this threshold for some entangled states. A simple example of such a
witness in the two-qubit case is a projector onto a Bell state with
the threshold at $\frac{1}{2}$.

All unit-trace positive operators can serve as statistical
operators. The largest mean value of a statistical
operator $\rho_{\mathrm{w}}$ is its largest eigenvalue
$\lambda_{\mathrm{max}}$, and if the largest mean value $\mu$ of
$\rho_{\mathrm{w}}$ with all separable states satisfies
$\mu<\lambda_{\mathrm{max}}$, then $\rho_{\mathrm{w}}$ is a
witness with the threshold at $\mu$.
Accordingly, $\rho_{\mathrm{w}}$ can serve as a
witness if its eigenspace to the largest
eigenvalue does not contain any product
state, and only then.  As a consequence, every entangled pure
state is a witness, and because almost all pure states are
entangled, it is easy to construct an IC-POM
with rank-one witnesses as outcomes.

To improve the witness efficiency (the probability of detecting a
random entangled state), it would be favorable to have optimal
witnesses \cite{Lewenstein2} as outcomes. Since rank-one POMs are
the best choice for tomographic purposes, it is of much concern
whether a pure-state witness can be an optimal witness. As we shall
see shortly, the efficiency of pure-state witnesses is poor, and
there are more efficient witnesses for the construction of WOMs.

First we  show that among the pure-state witnesses, the only candidates
for optimal witnesses are maximally entangled states in the case of
${d_1=d_2}$. The optimality of a pure-state witness is
determined by its Schmidt coefficients, and without loss of
generality we can assume
${\rho_{\mathrm{w}}=|\Psi\rangle\langle\Psi|}$,
${|\Psi\rangle=\sum_{i=1}^{d_1}|ii\rangle}\sqrt{\lambda_i }$ with the
Schmidt coefficients $\sqrt{\lambda_i}$ in nonincreasing order. The
eigenvalues of the partial transpose $\rho_{\mathrm{w}}^{T_2}$ are
\begin{equation}
\lambda_i\;( i=1,\cdots, d_1)\,,\  \pm
\sqrt{\lambda_i\lambda_j}\;(i,j=1,\cdots, d_1, i<j)\,.
\end{equation}
It follows that, for separable states $\rho_{\mathrm{sep}}$,
\begin{equation} \label{witnessThreshold}
\mathrm{Tr}\{\rho_{\mathrm{sep}}\rho_{\mathrm{w}}\}
=\mathrm{Tr}\bigl\{\rho_{\mathrm{sep}}^{T_2}\rho_{\mathrm{w}}^{T_2}\bigr\}\leq
\lambda_1\,,
\end{equation}
and $\rho_{\mathrm{sep}}=|11\rangle\langle 11|$ achieves this bound,
so the threshold of the pure-state witness $\rho_{\mathrm{w}}$ can
be set at $\lambda_1$. To determine the optimality of such a
witness, we construct the corresponding witness in the usual sense,
that is: with the threshold at zero,  $W=\lambda_1
\mathbf{1}-|\Psi\rangle\langle\Psi|$, where $\mathbf{1}$ denotes the
identity operator. Its partial transpose $Q=W^{T_2}$, which is a
positive operator, is
\begin{eqnarray}\label{Q}
Q&=&\sum_{i=1}^{d_1} |ii\rangle(\lambda_1-\lambda_i)\langle ii|
+\sum_{i,j=1,i<j}^{d_1}\Bigl(\lambda_1-\sqrt{\lambda_i\lambda_j}\Bigr)P_{ij}^{(+)}
\nonumber\\
&&+\sum_{i,j=1,i<j}^{d_1}\Bigl(\lambda_1+\sqrt{\lambda_i\lambda_j}\Bigr)
P_{ij}^{(-)}+\lambda_1P^{(\mathrm{r})}
\end{eqnarray}
with
\begin{eqnarray}
P_{ij}^{(\pm)}&=&
\frac{1}{2}\bigl(|ij\rangle\pm|ji\rangle\bigr)
\bigl(\langle ij|\pm\langle ji|\bigr)\,,\nonumber\\
P^{(\mathrm{r})}&=&\sum_{i=1}^{d_1}\sum_{j=d_1+1}^{d_2}|ij\rangle\langle ij|\,.
\end{eqnarray}
If $d_2>d_1$ or
$|\Psi\rangle$ is not maximally entangled, the range of $Q$ always
contains some product vector and, according to Theorem~2 of
\cite{Lewenstein2}, $W$ is not optimal. If $d_1=d_2$, and
$|\Psi\rangle$ is a maximally entangled state, then $Q$ is
proportional to the projector onto the antisymmetrical subspace
which contains no product vector in its range, and $W$ could be an
optimal witness.

The probability that a random pure  state is detected by a pure-state
witness  decreases exponentially with the dimension of
the Hilbert space.
To demonstrate this point we note that
pure states $|\Phi\rangle$ can be represented as points on a $2d_1d_2-1$
dimensional unit sphere (many points may correspond to the same state,
but it doesn't matter, as we are only concerned with the ratio), and
the set of pure product states is of measure zero.
The states detected by $\rho_{\mathrm{w}}$ form a spherical cap
determined  by
$\langle\Phi|\rho_{\mathrm{w}}|\Phi\rangle>\lambda_1$. The detection
probability is equal to the ratio of the hyper-area of the spherical
cap to that of the sphere, which is given by
\begin{equation}\label{RatioPure}
(1-\lambda_1)^{d_1d_2-1}\quad\mbox{with}\quad \lambda_1\geq1/d_1\,,
\end{equation}
where $\lambda_1$ is the square of the largest Schmidt coefficient
of $\rho_{\mathrm{w}}=|\Psi\rangle\langle\Psi|$. The maximum
detection probability of $(1-1/d_1)^{d_1d_2-1}$
is achieved when $|\Psi\rangle$ is maximally entangled. This
maximum detection probability approaches $e^{-d_2}$ for $d_1\gg1$.
So, even for the best pure-state witnesses, the
detection probability decreases exponentially with the dimension,
and it is desirable to find more efficient witnesses.

We restrict our attention to decomposable witnesses. An optimal
decomposable witness $W$ can be written as the partial transpose of
some positive operator $Q$ whose range contains no product vector
\cite{Lewenstein2}. We take a probabilistic approach in the study of
detection efficiency of a generic witness constructed from the
partial transpose of a statistical operator $Q$, see also
\cite{Znidaric}, because the method used in the calculation of
detection efficiency of a  pure-state witnesses is generally
difficult to carry out and may not give an intuitive result. When
$\Phi$ is distributed according to the normalized unitarily
invariant Haar measure $\mathrm{d}\mu(\Phi)$, the expectation value
and variance of the random variable
$\langle\Phi|Q^{T_2}|\Phi\rangle$ are
\begin{eqnarray}\label{MeanVar}
\mathrm{E}\bigl[\langle\Phi|Q^{T_2}|\Phi\rangle\bigr]&=&\frac{1}{d}\,,
\nonumber\\
\mathrm{Var}\bigl[\langle\Phi|Q^{T_2}|\Phi\rangle\bigr]&=&
\frac{\mathrm{Tr}\bigl\{(Q-\mathbf{1}/d)^2\bigr\}}{d(d+1)}.
\end{eqnarray}
We note that the variance is proportional to the squared
Hilbert--Schmidt (HS) distance between $Q$ and the maximally mixed
state.

In high dimension, there are generally many positive and negative
eigenvalues of $Q^{T_2}$ distributed randomly for a generic $Q$,
and as a consequence of the central-limit theorem, the
distribution of the random variable
$\langle\Phi|Q^{T_2}|\Phi\rangle$ will approximate a Gaussian
distribution, so that the probability of obtaining a negative value
is mainly determined by the ratio of the expectation value to the
standard deviation.

It would be favorable to maximize the standard deviation in order to
increase the detection efficiency. Recall that the expectation value
of the purity $\mathrm{Tr}\{Q^2\}$ with respect to the HS
measure and Bures measure are \cite{Karol, Ingemar}
\begin{equation}
\mathrm{E}\bigl[\mathrm{Tr}\bigl\{Q^2\bigr\}\bigr]_{\mathrm{HS}}
=\frac{2d}{d^2+1}\,,\quad
\mathrm{E}\bigl[\mathrm{Tr}\bigl\{Q^2\bigr\}\bigr]_\mathrm{B}
=\frac{5d^2+1}{2d(d^2+2)}\,,
\end{equation}
respectively, which both scale ${\propto 1/d}$ for ${d\gg1}$. So, if
we choose $Q$ randomly according to either of these two measures, or
other measures commonly used, the ratio of
$\left(\mathrm{E}\bigl[\langle\Phi|Q^{T_2}|\Phi\rangle\bigr]\right)^2$
to $\mathrm{Var}\bigl[\langle\Phi|Q^{T_2}|\Phi\rangle\bigr]$ is on
the order of $d$  with high probability, and the detection
probability decreases exponentially with growing $d$. It is now
clear why a pure-state witness can only detect a very small fraction
of entangled states in high dimension: because, when turned into a
witness $W$ in the usual sense, the positive operator $Q=W^{T_2}$ is
highly mixed, see Eq.~(\ref{Q}).

However, if $Q$ is a pure entangled state, the witness $W$ is
optimal \cite{Lewenstein3}, and the ratio is of order~1. Also,
in high dimension, a randomly chosen pure state is approximately a
maximally entangled state with high probability. For a maximally
entangled state $Q$, the detection probability of $W$ can be
calculated similarly to the case of pure-state witnesses, with the
result
\begin{eqnarray}
&&\frac{\int_0^{\pi/4}(\cos\alpha)^{d_1(d_1-1)-1}(\sin\alpha)^{d_1(d_1+1)-1}
\mathrm{d}\alpha}
{\int_0^{\pi/2}(\cos\alpha)^{d_1(d_1-1)-1}(\sin\alpha)^{d_1(d_1+1)-1}
\mathrm{d}\alpha}\nonumber\\
&=&\frac{\Gamma(d_1^2)}{2^{d_1^2}
\Gamma{\left(\frac{d_1(d_1+1)}{2}\right)}}\sum_{k=0}^{d_1}\frac{{d_1
\choose
k}(-1)^k\Gamma{\left(\frac{k+1}{2}\right)}}
{\Gamma{\left(\frac{d_1(d_1-1)+k+1}{2}\right)}}\,.
\end{eqnarray}
This ratio becomes
${(2\pi)^{-\frac{1}{2}}\int_1^{\infty}e^{-\frac{1}{2}x^2}\mathrm{d}x\approx0.1573}$
in the limit ${d_1\to\infty}$; see also \cite{Znidaric}. Numerical
calculation shows that when $d_1\geq7$, the deviation of the ratio
from the limit is within $1\%$, the largest deviation occurs when
${d_1=2}$, where the ratio is $\frac{1}{8}$. In conclusion, in high
dimension, a witness constructed from the partial transpose of a
randomly chosen pure state will, with high probability, detect about
$15.7\%$ of pure entangled states, thus achieving exponential
improvement over a pure-state witness.

We can now construct WOMs with optimal witnesses as outcomes. Given
a rank-one POM with outcomes $w_i\rho_i$, ${\sum_iw_i
\rho_i=\mathbf{1}}$, where $w_i>0$ and the $\rho_i$s are projectors
to pure entangled states, we can construct a WOM with outcomes
$\propto w_i\rho_{i\mathrm{w}}$, where
$\rho_{i\mathrm{w}}=\lambda_{\mathrm{max}}\mathbf{1}-\rho_i^{T_2}$
with $\lambda_{\mathrm{max}}$ the maximum of the largest eigenvalues
of $\rho_i^{T_2}$s. This WOM can achieve exponential improvement
with respect to witness efficiency, and if the POM is IC, so is the
WOM.

The outcomes of this WOM are highly mixed, and this decreases the
tomographic efficiency. To compare the tomographic efficiencies of
the WOM and the POM, we consider an example in which the POM is a
symmetric IC-POM (SIC-POM), which was conjectured to exist in finite
dimensions \cite{Zauner, Renes, Appleby}. The mean square errors
(according to the HS distance) achieved by the SIC-POM and the
optimal WOM are given by \cite{Scott,Zhu1}
\begin{eqnarray}
&&\mathrm{E}\bigl(|\!|\hat{\rho}-\rho|\!|_{\mathrm{HS}}^2\bigr)_{\mathrm{POM}}
=\frac{1}{N}{\left(d^2+d-1-\mathrm{Tr}\bigl\{\rho^2\bigr\}\right)}
\sim\frac{d^2}{N}\,,\nonumber\\
&&\mathrm{E}\bigl(|\!|\hat{\rho}-\rho|\!|_{\mathrm{HS}}^2\bigr)_{\mathrm{WOM}}
=\frac{1}{N}\biggl(\frac{1}{d}\bigl[(d+1)^2(d-1)\nonumber\\
&&\quad \times(d\lambda_{\mathrm{max}}-1)^2
+1\bigr]-\mathrm{Tr}\bigl\{\rho^2\bigr\}\biggr)
\sim\frac{\lambda_{\mathrm{max}}^2d^4}{N}\,,\label{MSE}
\end{eqnarray}
where $\hat{\rho}$ is an estimate of the input state $\rho$ in
accordance with the measurement results, and $N$ is the number of
copies of the input state available for state tomography. Note that
$\lambda_{\mathrm{max}}\geq 1/d_1$ (if every fiducial vector of the
SIC-POM is approximately maximally entangled,
$\lambda_{\mathrm{max}}\sim1/d_1$), so in high dimension, to achieve
the same mean square error, the number of copies of an input state
required in the WOM is at least $d_2^2$ times larger. For
different rank-one POMs, the specific reduction of the tomographic
efficiency may be different, but the order of magnitude should be
similar. The trade-off between the tomographic
efficiency and the witness efficiency is not restricted to this
specific example, as we have seen that rank-one outcomes, which are
best for tomographic purposes, are generally very poor as
witnesses, the more so in high dimension. Generally
speaking, positive operators with high purity tend to be poor
witnesses, and those with low purity are not desirable
for tomography.

Detection of  entanglement of mixed states turns out to be much more
difficult \cite{Znidaric}, even if $Q$ is a pure state. For
${W=\bigl(|\Psi\rangle\langle\Psi|\bigr)^{T_2}}$,
the distribution of the random
variable  ${\mathrm{Tr}\{\rho W\}=\langle\Psi|\rho^{T_2}|\Psi\rangle}$,
is the same in form as that of the random variable
$\langle\Phi|Q^{T_2}|\Phi\rangle$ discussed above, provided $\rho$
varies among all unitarily equivalent statistical operators
according to the Haar measure. In particular, the expectation value
and variance of the random variable is the same as in
Eq.~(\ref{MeanVar}) after  replacing $Q$ by $\rho$. As the fraction
of separable states approaches zero in high dimension, by similar
arguments as before, we can conclude that the probability of
detecting entanglement of a random mixed state will generally
decrease exponentially as the purity of the input state decreases.

Despite exponential improvement with respect to witness efficiency,
the WOM with optimal witnesses as outcomes is still not efficient
enough to detect highly mixed entangled states. In that case,
entanglement detection through state tomography may be a better
choice. To estimate the number of copies required in state
tomography to reach sufficient accuracy for determining the
separability of an input state, we recall that the radius of the
largest separable ball in the space of statistical operators is
${\sqrt{1/(d-1)d}\approx 1/d}$ \cite{Gurvits}, the dimension of the
state space is $d^2-1\approx d^2$, and a reasonable estimate of
accuracy requirement would be
${\mathrm{E}\bigl(|\!|\hat{\rho}-\rho|\!|_{\mathrm{HS}}\bigr)\sim1/d^3}$,
${\mathrm{E}\bigl(|\!|\hat{\rho}-\rho|\!|_{\mathrm{HS}}^2\bigr)\sim
1/d^6}$; for state tomography with a SIC-POM this would mean ${N\sim
d^8}$.

For illustration, we now turn to the situation of two qubits, and
analyze the difference between rank-one POMs
and WOMs with optimal witnesses.
In the case of two qubits, there are only decomposable
witnesses \cite{Lewenstein2}. Since every rank-two subspace
contains at least one product vector \cite{Hill, Englert2}, the
only candidates of optimal witnesses (in the usual
sense) are the partial transposes of some entangled pure states, and
they are indeed optimal \cite{Lewenstein2}.
We first compare pure-state witnesses and
witnesses constructed from their partial transposes.

By some local unitary transformation, any entangled two-qubit state
can be turned into the form,
\begin{eqnarray}\label{StandardForm}
&|\Psi\rangle=|00\rangle\cos\alpha+|11\rangle\sin\alpha\,,
&\nonumber\\
&\displaystyle 0<\alpha
=\frac{1}{2}\arcsin(q)
\leq\frac{\pi}{4}\,,\quad p=\sqrt{1-q^2}\,,&
\end{eqnarray}
where $q$ is the concurrence of the pure state, $p$ is the length of
the Bloch vector of each reduced density matrix, and $\alpha$ is
introduced for convenience. Because either one of the parameters
$q,p,\alpha$ specifies $|\Psi\rangle$, we will use them
interchangeably to denote a pure state with given concurrence. Two
pure states are \emph{aligned} if there exists a local unitary
transformation which turns both of them into the form of
Eq.~(\ref{StandardForm}), possibly with  different parameters
$\alpha_1, \alpha_2$ or, equivalently, if the Bloch vectors of their
reduced statistical operators are parallel, and the fidelity between
them is equal to that between their reduced statistical operators.
The fidelity between two  pure states with given concurrences
obtains the maximum value of $\cos(\alpha_1-\alpha_2)^2$ when they
are aligned.

If ${\alpha<\frac{\pi}{4}}$, the witness constructed from the pure
state is not optimal, and Lemma 2 of \cite{Lewenstein2} implies that
the witnesses constructed from its aligned pure states with larger
concurrences are finer. The probability that a random pure entangled
state is detected by a pure-state witness with concurrence $q$ is
$\bigl[\frac{1}{2}(1-p)\bigr]^3$ as follows from
Eq.~(\ref{RatioPure}), with the maximum $\frac{1}{8}$ achieved for a
Bell state. For two aligned pure states $\rho(\alpha_1)$ and
$\rho(\alpha_2)$, state $\rho(\alpha_2)$ can detect state
$\rho(\alpha_1)$ if
\begin{equation}
0<\mathrm{Tr}\{\rho(\alpha_1)\rho(\alpha_2)\}-
\cos(\alpha_2)^2=\sin(\alpha_1)\sin(2\alpha_2-\alpha_1)\,,
\label{DetectCon}
\end{equation}
which means  ${2\alpha_2>\alpha_1}$. A counterintuitive consequence of
this observation is that, when ${\alpha\leq\frac{\pi}8}$  or
${q\leq\frac{1}{2}\sqrt{2}}$, $\rho(\alpha)$ cannot detect any Bell
states. An analogous phenomenon also exists in high dimension: if
one Schmidt coefficient of an entangled pure state is particularly
large, then it cannot detect any maximally entangled states.

The four eigenvalues of $\rho(\alpha)^{T_2}$ are $\frac{1}{2}(1\pm p)$,
$\pm\frac{1}{2}q$ \cite{Englert2}, so that statistical operators,
which are also optimal witnesses related to $\rho(\alpha)$,
can be constructed as
\begin{equation}
\rho_{\mathrm{w}}(\alpha)
=\frac{\frac{1+p}{2}\mathbf{1}-\rho(\alpha)^{T_2}}{1+2p}\,.
\label{WitnessMap}
\end{equation}
Such a $\rho_{\mathrm{w}}(\alpha)$ has concurrence ${q/(1+2p)}$ and negativity
${(1-p)/(1+2p)}$, and ${\mu=(1+p)/(2+4p)}$
is its witness threshold.
These states are quite special, they are Bell states if
${\alpha=\frac{\pi}{4}}$, and are maximally entangled mixed states for other
values of $\alpha$, those states whose concurrence cannot be increased by
any global unitary transformation \cite{Verstraete}.
When  $\rho_{\mathrm{w}}(\alpha_1),
\rho_{\mathrm{w}}(\alpha_2)$ are aligned --- we call
$\rho_{\mathrm{w}}(\alpha_1), \rho_{\mathrm{w}}(\alpha_2)$ aligned
if $\rho(\alpha_1),\rho(\alpha_2)$ are aligned ---
$\rho_{\mathrm{w}}(\alpha_2)$ can detect $\rho_{\mathrm{w}}(\alpha_1)$ if
\begin{equation}
0>\mathrm{Tr}\Bigl\{\rho_{\mathrm{w}}(\alpha_1){\rho(\alpha_2)}^{T_2}\Bigr\}
=\frac{\sin(\alpha_2)\sin(\alpha_2-2\alpha_1)}{1+2p_1}\,,
\end{equation}
which means ${2\alpha_1>\alpha_2}$, a condition that is dual to
the condition of Eq.~(\ref{DetectCon}). If $\alpha_1>\frac{\pi}{8}$, so that
$\rho_{\mathrm{w}}(\alpha_1)$ could be any Bell state, then
$\rho_{\mathrm{w}}(\alpha_1)$ can be detected by any aligned optimal
witness $\rho_{\mathrm{w}}(\alpha_2)$. On the other hand, an
optimal witness $\rho_{\mathrm{w}}(\alpha_2)$ cannot
detect any state $\rho_{\mathrm{w}}(\alpha_1)$ with
$\alpha_1\leq\alpha_2/2$; in particular, Bell states cannot detect
any states $\rho_{\mathrm{w}}(\alpha_1)$ with $\alpha_1\leq\frac{\pi}{8}$.
As $\alpha,q\rightarrow0$, the
$\rho_{\mathrm{w}}(\alpha)$s approach separable states, and they can
detect fewer and fewer entangled states but, surprisingly, they
are still the best witnesses for detecting even more weakly entangled
states $\rho_{\mathrm{w}}(\alpha^\prime)$ with
$\alpha^\prime\leq\alpha$. As a consequence, an infinite number of
witnesses are needed to detect all entangled states, if
we rely on each witness separately.

Taking a SIC-POM \cite{Zauner, Renes, Appleby} as example, we shall
now show the sharp difference between a rank-one POM and a WOM
consisting of optimal witnesses. Most known examples of SIC-POMs are
constructed from fiducial states under the action of the generalized
Pauli group, or Heisenberg--Weyl group. In four dimensions, one such
fiducial state is given in Eq.~(147) of \cite{Appleby}. In the case of
two qubits, if we choose the standard product basis
$|00\rangle,|01\rangle,|10\rangle,|11\rangle$ for Appleby's
 $|e_0\rangle,|e_1\rangle,|e_2\rangle,|e_3\rangle$, then
the SIC-POM thus constructed has the very peculiar property that the
concurrence of all fiducial states are the same, namely
$\sqrt{2/5}$~\cite{Zhu2}.
We note that these fiducial states are typical in the sense that their squared
concurrence equals the average squared concurrence of all pure two-qubit
states~\cite{Karol}. 

\begin{figure}
  \centerline{\includegraphics{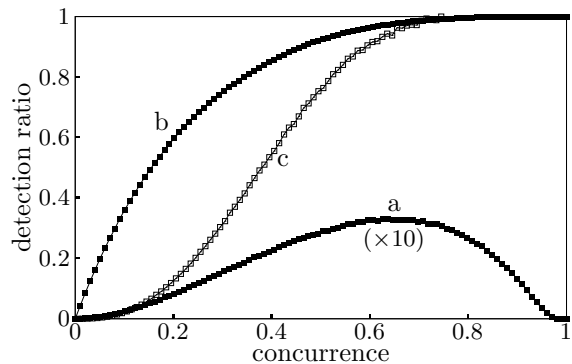}}
\caption{\label{fig:ratio}%
Simulation results for the entanglement detection ratio of the SIC-POM (curve~a)
and the optimal WOM (curves b and~c), both for pure states (curves a and~b)
and mixed states (curve~c), in the case of two qubits.
The detection ratio of the SIC-POM (curve~a) has been multiplied by
$10$ for ease of viewing; note that the SIC-POM does not detect states with a
concurrence close to $1$.
For curve c, $10^7$ mixed states were generated randomly
according to the HS measure,
applying a procedure described in Ref.~\cite{Karol};
the SIC-POM gave no counts for mixed states.
Curve c stops at concurrence $0.75$ because there are too few counts for
reliable statistics for larger concurrence values.}
\end{figure}

According to the general procedure presented above, we can
construct a WOM with optimal witnesses as outcomes.
Moreover, the WOM is literally also SIC,
except that the outcomes are not pure.
Figure~\ref{fig:ratio} shows the entanglement detection ratio for
the SIC-POM and the WOM for both pure and mixed states, with
each witness acting separately. There is a huge improvement of the
WOM over the SIC-POM with respect to  witness efficiency
especially for the detection of mixed states. The SIC-POM cannot
detect any Bell states, because the concurrence of each fiducial
vector is less than $\sqrt{1/2}$; yet, the detection ratio of the
WOM approaches 1 as the concurrence of the input state
approaches 1. If mixed states are distributed according to the
HS measure, the overall detection probability of the
WOM is about 13\%, for states with concurrence larger than $\frac{1}{2}$,
the probability is about 84\%.

The improvement in the witness efficiency comes at the price of a
much lower tomographic efficiency, the mean square error achieved by
the SIC-POM and the WOM given in Eq.~(\ref{MSE}) now reads,
\begin{eqnarray}
\mathrm{E}\bigl(|\!|\hat{\rho}-\rho|\!|_{\mathrm{HS}}^2\bigr)_{\mathrm{POM}}
&=&\frac{1}{N}{\left(19-\mathrm{Tr}(\rho^2)\right)},\nonumber\\
\mathrm{E}\bigr(|\!|\hat{\rho}-\rho|\!|_{\mathrm{HS}}^2\bigr)_{\mathrm{WOM}}
&=&\frac{1}{N}\bigl(64+15\sqrt{15}-\mathrm{Tr}\{\rho^2\}\bigr)\,,\quad
\end{eqnarray}
where we have inserted the value
${\lambda_{\mathrm{max}}=\frac{1}{2}(1+\sqrt{3/5})}$. The reduction in the
tomographic efficiency of the WOM compared to the
SIC-POM is roughly by a factor of~$\frac{2}{13}$.

In summary, we have introduced WOMs, and
developed a systematic method of constructing a WOM with
optimal witnesses as outcomes from a rank-one POM.
This WOM can provide exponential improvement with respect to
the witness efficiency in high dimension, at the price of a
reduction in the tomographic efficiency by a factor of order $d_2^{-2}$.
The trade-off between the witness efficiency and the
tomographic efficiency is quite common --- this reflects the intrinsic
difficulty of entanglement detection.

We are grateful for valuable discussions with Artur \mbox{Ekert}
and Karol \.Zyczkowski. 
HJZ thanks  Maciej Lewenstein and Lin Chen for assistance and advice.
Centre for Quantum Technologies is a Research Centre of Excellence
funded by Ministry of Education and National Research Foundation of
Singapore.


\begin{thebibliography}{99}
\bibitem{PO(V)M}
POM, with its emphasis on ``probability'' and ``measurement'' is quantum physics
jargon.
The corresponding popular mathematical term POVM
(positive-operator-valued measure) makes reference to measure theory.
\end{thebibliography}
\end{document}